\documentclass[9pt, twocolumn]{revtex4}
\pdfoutput=1
\usepackage{hyperref,amsmath, mhchem}
\usepackage{bbold,mathrsfs}

\newcommand{\dg}{^\dagger}
\newcommand{\hf}{\frac{1}{2}}

\newcommand{\be}{\begin{equation}}
\newcommand{\ee}{\end{equation}}
\newcommand{\bes}[2]{\begin{equation}\begin{split}#1\end{split}\label{#2}\end{equation}}

\usepackage{tikz}
\usetikzlibrary{calc}
\usetikzlibrary{arrows}
\usetikzlibrary{decorations.markings}
\usepackage{pgfplots}
\usepackage{rotating}
\usepackage{graphicx}
\usetikzlibrary{pgfplots.groupplots}
\tikzset{
    axis break gap/.initial=0mm
}
\usepgfplotslibrary{fillbetween}

\usepackage{multirow}
\usepackage{booktabs, ctable}
\usepackage{array}
\newcolumntype{M}[1]{>{\centering\arraybackslash}m{#1}}
\usepackage{soul}

\usepackage{adjustbox}

\begin{document}
\definecolor{mybl}{rgb}{0.00000,0.44700,0.74100}

\title{Vortex Lattices in Binary Mixtures of Repulsive Superfluids}

\date{\today}
\author{Luca Mingarelli,  Eric E Keaveny,  and Ryan Barnett}
\affiliation{Department of Mathematics, Imperial College London,
London SW7 2AZ, United Kingdom}
\begin{abstract}
We present an extension of the framework introduced in \cite{Noi} to treat multicomponent systems, showing that new degrees of freedom are necessary in order to obtain the desired boundary conditions. We then apply this extended framework to the coupled Gross-Pitaevskii equations to investigate the ground states of two-component systems with equal masses 
thereby extending previous work in the lowest Landau limit
\cite{Mueller02} to arbitrary interactions within Gross-Pitaevskii theory.
We show that away from the lowest-Landau level limit, the predominant vortex lattice consists of two interlaced triangular lattices. 
Finally, we derive a linear relation which accurately describes the phase boundaries in the strong interacting regimes.

\end{abstract}
\maketitle

\section{Introduction}
The way a superfluid acquires angular momentum is perhaps one of its most interesting properties. 
Ever since the experimental verification \cite{Farnoux64,Essmann67} of early predictions \cite{Abrikosov57,Kleiner64,Tkachenko66} of a ground state consisting of a lattice of singly quantised vortices, many advances have been made in the field, and
the nucleation of vortices carrying quantised circulation
has attracted much interest \cite{Onsager49,Feynman55_2,Donnelly,Sonin87,Fetter09}. 
Bose-Einstein condensates have proven particularly suitable to study vortex lattices in superfluids: while in the early experiments it was only possible to create a few vortices, it is now possible to obtain and study condensates with over $100$ vortices in systems with a lifetime of several seconds \cite{Abo-Shaeer476}.

While the behaviour of a single component superfluid is today a more and more understood problem, the same is not true for the next simplest case of two interacting superfluids. The behaviour of such systems has been the subject of study in both their attractive and repulsive regimes \cite{Barnett10,Kuopanportti12,Mason11,Kasamatsu03,Keifmmode06}, although classification has proved difficult due to various challenges (e.g.~lattice distortion from the trap). In particular in \cite{Mueller02}, the vortex lattice configurations in repulsive two-component superfluids with equal masses and equal intra-component interactions were found in the limit of fast rotation. In this limit, the condensate order parameter can be projected into the lowest Landau level (LLL) basis, which simplifies the analysis. In this work we extend the results of \cite{Mueller02} to arbitrary rotation rates, and hence explore the full phase diagram of experimentally accessible regimes. We find that the vortex lattice configurations predicted in \cite{Mueller02} survive away from the LLL regime. However, for slow rotation rates (or strong interactions) we find that the phase diagram becomes dominated by the triangular lattice configuration. The main result of this work is summarised in the phase diagram of Fig.~\ref{fig:phasespace}.

{Our analysis relies on the computational framework described in \cite{Noi} for the case of a single component superfluid. There, the lowest energy solution to the Gross-Pitaevskii equation was obtained in a quasi-periodic unit cell by means of the so-called \emph{Magnetic Fourier transform} which gives a straightforward diagonalisation of the relevant linear operators of the model, therefore allowing for efficient solutions.}
The removal of the distortion of the vortex lattice caused by the trap \cite{Sheehy04} allows for the exact characterisation of each configuration in terms of parameters directly entering the Gross-Pitaevskii energy functional. We can therefore directly explore the phase transitions occurring in different inter and intra-species interaction regimes. The generalisation of the framework \cite{Noi} to multi-component systems is not immediate. As a second main result, this work will describe how this generalisation is achieved.

For sake of generality and clarity, the intraspecies interaction of a superfluid with atomic mass $m_j$ can be quantified in terms of the dimensionless ratios of two characteristic lengths of the system $\ell_B^{(j)}/\xi^{(j)}$. The healing lengths $\xi^{(j)}=\sqrt{\frac{\hbar^2}{2m_jg_j\bar{\rho}_j}}$, where $\bar{\rho}_j$ is the average superfluid density of the $j$th component and $g_j$ the $j$th intraspecies interaction strength, provide a measure of the characteristic core sizes of the vortices of each component; the magnetic lengths $\ell_B^{(j)}=\sqrt{\frac{\hbar}{2\Omega m_j}}$, introduced in analogy with quantum Hall systems \cite{Stone92}, give instead a measure of the characteristic separation between vortices with $\Omega$ being the rotational frequency of the system. In the following we will be particularly concerned with the case in which such a dimensionless ratio is the same for both components: $\ell_B/\xi\equiv\ell_B^{(1)}/\xi^{(1)}=\ell_B^{(2)}/\xi^{(2)}$.  We further restrict to equal average densities $\bar{\rho}_1 = \bar{\rho}_2$.  
We impose these restrictions for simplicity and note that
no qualitative aspects of the 
conclusions reached in this work are affected by  lifting them.
In order to characterise the inter-component interaction instead,
we introduce the dimensionless quantity $\alpha=\frac{g_{12}}{\sqrt{g_1g_2}}$ \cite{Mueller02}, where $g_{12}$ is the inter-component interaction strength, which allows us to consider the interaction between the two species relative to their intra-component interactions rather than as an absolute quantity. This is important as the boundary $\alpha=1$ corresponds to the condition for the two species to be miscible or immiscible \cite{Pethick}. 

This paper is organised as follows. In sections \ref{Sec:2} and \ref{Sec:3} we discuss the mean-field Gross-Pitaevskii theory describing the system and extend the procedure introduced in \cite{Noi} to the multicomponent case. In the following section \ref{Sec:4} we extend the results of \cite{Mueller02} providing a detailed investigation of the phase space spanned by $\alpha$ and $\ell_B/\xi$. In what follows, we will refer to the limit opposite to the LLL, i.e. the limit of strong intraspecies interactions, as the Coulomb limit. In this regime, one can treat the intra-species interactions with a Coulomb-like potential to a good approximation.

\section{The multicomponent case}
\label{Sec:2}
The two-dimensional energy functional associated with a two-species system in a rotating frame of reference is given, within Gross-Pitaevskii mean field theory, by 
\bes{E=\int\mathcal{E}[\psi_1,\psi_2]\text{d} x\text{d} y,}{GPEnergy}
where the energy density is
\bes{
\mathcal{E}[\psi_1,\psi_2]=&\sum_{j=1}^2 \Bigl[\frac{\hbar}{2m_j}|\nabla\psi_j|^2+\hf m_j\omega_j^2r^2|\psi_j|^2\\
&-\psi_j\dg\Omega L_z\psi_j-\mu_j|\psi_j|^2\Bigr]+\hf{\boldsymbol{\rho}^T \mathcal{G}\boldsymbol{\rho}}.
}{GPEnergydens}
Here, $L_z=-i\hbar(x\partial_y-y\partial_x)$ is the angular momentum operator along the $z$-axes, $\Omega$ is the rotational frequency, and $m_j$, $\mu_j$, $\omega_j$ are respectively the mass, the chemical potential, and the trapping frequency of the $j$th species. The matrix
\bes{
\mathcal{G}=\begin{pmatrix}
    g_{1} & g_{12} \\
    g_{12}  & g_2 \\
    \end{pmatrix}
}{}
accounts for intra and inter-species interactions which are related to the s-wave scattering lengths $a_{jk}$: $g_j=4\pi\hbar^2a_{jj}/m_j$,
$g_{12}=4\pi\hbar^2a_{12}(m_1+m_2)/m_1m_2$. Finally $\boldsymbol{\rho}^T=(|\psi_1|^2,|\psi_2|^2)$. The miscibility condition which ensures the two species do not phase separate is for $\mathcal{G}$ to be positive semi-definite; this can be analogously expressed in terms of the dimensionless parameter previously introduced $\alpha\le1$. 

The energy density functional \eqref{GPEnergy} can be rearranged in a convenient way: introducing the symmetric gauges $\boldsymbol{A}_j=\Omega m_j(-y,x)$ and setting the effective frequencies $\omega^{\text{eff}}_j=\sqrt{\omega_j^2-\Omega^2}=0$, we can write \eqref{GPEnergy} as
\bes{
\mathcal{E}[\psi_1,\psi_2]&=\sum_{j=1}^2 \left[\frac{1}{2m_j}\left|(-i\hbar\nabla-\boldsymbol{A}_j)\psi_j\right|^2-\mu_j|\psi_j|^2\right]\\
&+\hf{\boldsymbol{\rho}^T \mathcal{G}\boldsymbol{\rho}} .
}{energydens}
This leads to the two corresponding coupled Gross-Pitaevskii equations $i\hbar\partial_t\psi_j=\delta E/\delta\psi_j^*$ describing the dynamics of the system. The above form of the energy density functional is particularly appealing as it makes the gauge invariance of the system explicit. This property 
allows us to switch to the Landau gauge ${\bf A}_j^{\bf L}=2\Omega m_jx\hat{y}$ without affecting the energy functional \eqref{GPEnergy}. Such a perspective will prove useful later.

Finally, let us comment on the allowed boundary conditions of such a system. The presence of a gauge field makes the standard periodic boundary conditions very unnatural and very large unit cell sizes need to be taken in practice. One can instead find the wavefunction satisfies twisted boundary conditions, consisting of the acquisition of a phase when moving over a period. The twisted boundary conditions can be taken into account by employing the magnetic Fourier transform (MFT) which correctly diagonalises the linear part of the energy functional \cite{Noi}.

\section{Computational Framework}
\label{Sec:3}

In this section we describe the computational method used to find the minima of the two-component energy functional given in \eqref{GPEnergy}. This involves a non-trivial extension of the method described in \cite{Noi} which treats the single-component system.
We approach the problem of the discretisation of the energy \eqref{energydens} following \cite{Noi}. More specifically, upon defining $\Psi=(\psi_1,\psi_2)^T$, we consider the coupled non-linear Hofstadter model 

\bes{
E_{\text{d}}=&-\sum_{n,m}\Bigl[\Psi_{n,m}\dg\boldsymbol{W}^{(x)}
\boldsymbol{\Phi}^{(x)}_m
\Psi_{n+1,m}\\
&\phantom{\sum_{n,m}\Bigl[}+\Psi_{n,m}\dg\boldsymbol{W}^{(y)}\boldsymbol{\Phi}
^{(y)}_n
\Psi_{n,m+1}+\text{h.c.}\Bigr]\\
&+\sum_{n,m}\left[\hf\boldsymbol{\rho}_{n,m}\dg\mathcal{U}\boldsymbol{\rho}_{n,m}-\Psi_{n,m}\dg\overline{\boldsymbol{\mu}}\Psi_{n,m}\right],
}{discrete_energy}
defined on a grid of ${N_x\times N_y}$ points taking values ${{\bf r}=a_xn\hat{\bf x}+a_ym\hat{\bf y}}$, with $n,m\in\Bbb{Z}^+$, $n\le N_x$, $m\le N_y$, with lattice constants $a_k=L_k/N_k$, and with $L_x,L_y$ being the lengths of the computational unit cell. In the above, $\boldsymbol{W}^{(x)}$, $\boldsymbol{W}^{(y)}$ account for the anisotropic tunnelling for each component while the $\boldsymbol{\Phi}^{(k)}_n$  arise from the Peierls substitution \cite{Hofstadter76,Peierls33} needed to incorporate the gauge fields:
\bes{
\boldsymbol{W}^{(k)}&=\frac{\hbar^2}{2a_k^2}\begin{pmatrix}
    \frac{1}{m_1} & 0 \\
    0  & \frac{1}{m_2} \\
    \end{pmatrix},\\
        \boldsymbol{\Phi}^{(k)}_n&=\begin{pmatrix}
    e^{-i\mathcal{B}_1^{(k)}n} & 0 \\
    0  & e^{-i\mathcal{B}_2^{(k)}n} \\
    \end{pmatrix}
.
}{Wphimatrix}
It is well known that the discrete energy \eqref{discrete_energy} reduces to the energy functional \eqref{GPEnergy} provided that the lattice constant is the smallest length scale in the problem. In doing so, provided one considers the Landau gauge, it is possible to  verify the following identifications: ${\mathcal{B}_j^{(k)}=2\delta_{yk}\Omega m_ja_xa_y/\hbar}$, $\mathcal{U}=\mathcal{G}/a_xa_y$ and ${\overline{\boldsymbol{\mu}}=\text{diag}[\mu_1,\mu_2]-2(\boldsymbol{W}^{(x)}+\boldsymbol{W}^{(y)})}$.  Equivalently, an alternative to fixing the chemical potential is to fix the total particle numbers per unit cell as $\int|\psi_j|^2\text{d}x\text{d}y=\mathcal{N}_j$.

We next perform a local gauge transformation on the second component:
\bes{
\Psi_{n,m} &\rightarrow\begin{pmatrix}
    1 & 0 \\
    0  & e^{-i\lambda_{n,m}} \\
    \end{pmatrix}\Psi_{n,m},
}{phgt}
where the pure gauge is $\lambda=\tau_xa_xn+\tau_ya_ym$. 
Inserting this into (\ref{discrete_energy}), one finds that 
\bes{
\boldsymbol{\Phi}^{(k)}_n&\rightarrow\begin{pmatrix}
    1 & 0 \\
    0  & e^{-i\tau_ka_k}\\
    \end{pmatrix}\boldsymbol{\Phi}^{(k)}_n.    
}{phgt2}
A comment on the need for this gauge transformation will be given below. We further assume that $\Psi_{n,m}$ can be expanded in the basis of states ${\widetilde{\Psi}_{k_x,m}=\left(\widetilde{\psi}_{1;k_xm},\widetilde{\psi}_{2;k_xm}\right)^T}$ and ${\widetilde{\Psi}_{n,k_y}=\left(\widetilde{\psi}_{1;nk_y},\widetilde{\psi}_{2;nk_y}\right)^T}$ as
\bes{
\psi_{j;nm}=&
\frac{1}{\sqrt{N_x}}\sum_{k_x} e^{i(k_xn+\mathcal{B}_jnm)}\widetilde{\psi}_{j;k_xm},\\
\psi_{j;nm}=&
\frac{1}{\sqrt{N_y}}\sum_{k_y} e^{ik_ym}\widetilde{\psi}_{j;nk_y}.
}{iMFT}
This is equivalent to demanding $\Psi_{n,m}$ to be an eigenfunction of the magnetic translation operators  with eigenvalue equal to one. In doing so, we also automatically satisfy the required twisted boundary conditions \cite{Noi,Yang61}. Inverting the relation in \eqref{iMFT}, we can then define the discrete magnetic Fourier transform (dMFT) of the $j$th component as
\bes{
\widetilde{\psi}_{j;k_xm}=&
\frac{1}{\sqrt{N_x}}\sum_n e^{-i(k_xn+\mathcal{B}_jnm)}\psi_{j;nm},\\
\widetilde{\psi}_{j;nk_y}=&
\frac{1}{\sqrt{N_y}}\sum_m e^{-ik_ym}\psi_{j;nm},
}{MFT}
which will be fundamental for the diagonalisation of the problem at hand. 

A comment  is needed concerning the gauge transformation given above and the boundary conditions of the system.
The gauge transformation \eqref{phgt} has the effect of introducing two new degrees of freedom contributing to an overall phase of the second component's wavefunction. In \cite{Noi}, the wavefunctions were taken to be invariant when magnetically translated along a vortex lattice vector. While this constraint is appropriate for the single-component case, it must be relaxed for the multi-component system. For the present case, we must consider the whole set of possible states obtainable by translating one component with respect to the other. Clearly one needs to translate only one of the two components to obtain such a set. As described with further detail in the Appendix, the appropriate way to perform such translations is to employ an operator of the magnetic translation group \cite{Noi,Stone92}. Such a translation is accounted for by the parameters $(\tau_x, \tau_y)$ introduced in \eqref{phgt} and \eqref{phgt2}, as is also explained
in the Appendix.

As discussed in \cite{Noi}, the employment of the magnetic Fourier transform (MFT) diagonalises the kinetic part of the model. The expansion \eqref{iMFT} is of great importance as it allows, through its inverse \eqref{MFT}, for the diagonalisation of the linear (kinetic) part of the model \eqref{discrete_energy}.
The discrete energy \eqref{discrete_energy} can now be written compactly as\\
\bes{
E_{\text{d}}=&4\mathcal{R}\sum_{k_x,m}\widetilde{\Psi}_{k_x,m}\dg\overline{\boldsymbol{W}}\boldsymbol{K}^{(x)}_{k_x,m}\widetilde{\Psi}_{k_x,m}\\
&+\frac{4}{\mathcal{R}}\sum_{n,k_y}\widetilde{\Psi}_{n,k_y}\dg\overline{\boldsymbol{W}}\boldsymbol{K}^{(y)}_{n,k_y}\widetilde{\Psi}_{n,k_y}\\
&+\sum_{n,m}\left[\hf\boldsymbol{\rho}_{n,m}\dg\mathcal{U}\boldsymbol{\rho}_{n,m}-\Psi_{n,m}\dg\overline{\boldsymbol{\mu}}\Psi_{n,m}\right],
}{Ed}
where we have defined the matrices accounting for the kinetic terms
\bes{
\boldsymbol{K}^{(x)}_{k_x,m}=&\text{diag}\left[
    \sin^2\left(\frac{k_x+\mathcal{B}_1m}{2}\right), \sin^2\left(\frac{k_x+\mathcal{B}_2m+\tau_x}{2}\right) \right],\\
\boldsymbol{K}^{(y)}_{n,k_y}=&\text{diag}\left[
    \sin^2\left(\frac{k_y-\mathcal{B}_1n}{2}\right), \sin^2\left(\frac{k_y-\mathcal{B}_2n+\tau_y}{2}\right) \right],
}{}
and introduced ${\overline{\boldsymbol{W}}=\left(\boldsymbol{W}^{(x)}\boldsymbol{W}^{(y)}\right)^{\circ\hf}}$, denoting by `$\circ$' 
element-wise exponentiation. We have also introduced the aspect ratio $\mathcal{R}=\frac{L_y}{L_x}=\frac{a_y}{a_x}$, which explicitly accounts for anisotropic tunnelling.

Each term in \eqref{Ed} is now diagonal and the minimisation of the energy functional 
with respect to $\Psi$ can thus be achieved by solving the associated equations of motion in imaginary time in conjunction with a split-step method \cite{Noi}; a further minimisation is then required with respect to $\tau_x$, $\tau_y$ and $\mathcal{R}$. Holding $\widetilde{\Psi}$, $\tau_x$ and $\tau_y$ fixed, it is straightforward to show that \eqref{Ed} is minimised by requiring
\bes{
\mathcal{R}^*=\sqrt{\frac{\sum_{n,k_y}\widetilde{\Psi}_{n,k_y}\dg\overline{\boldsymbol{W}}\boldsymbol{K}^{(y)}_{n,k_y}\widetilde{\Psi}_{n,k_y}}{\sum_{k_x,m}\widetilde{\Psi}_{k_x,m}\dg\overline{\boldsymbol{W}}\boldsymbol{K}^{(x)}_{k_x,m}\widetilde{\Psi}_{k_x,m}}}.
}{minR}
Similarly one can find that holding $\widetilde{\Psi}$ and $\mathcal{R}$ fixed, the discrete energy \eqref{Ed} is minimised with respect to $\tau_x$ and $\tau_y$ by choosing
\bes{
\tau_x^*=-\arctan\left[\frac{\sum_{k_x,m}\sin(k_x+{\cal B}_2m)|\widetilde{\psi}_{k_x,m}^{(2)}|^2}{\sum_{k_x,m}\cos(k_x+{\cal B}_2m)|\widetilde{\psi}_{k_x,m}^{(2)}|^2}\right]\\+\pi \Theta\left(-\sum_{k_x,m}\cos(k_x+{\cal B}_2m)|\widetilde{\psi}_{k_x,m}^{(2)}|^2\right),\\
\tau_y^*=-\arctan\left[\frac{\sum_{n,k_y}\sin(k_y-{\cal B}_2n)|\widetilde{\psi}_{n,k_y}^{(2)}|^2}{\sum_{n,k_y}\cos(k_y-{\cal B}_2n)|\widetilde{\psi}_{n,k_y}^{(2)}|^2}\right]\\+\pi \Theta\left(-\sum_{n,k_y}\cos(k_y-{\cal B}_2n)|\widetilde{\psi}_{n,k_y}^{(2)}|^2\right),
}{alphabetastar}
where $\Theta(x)$ is the Heaviside function. \begin{figure}[h!]
	\centering
	\includegraphics{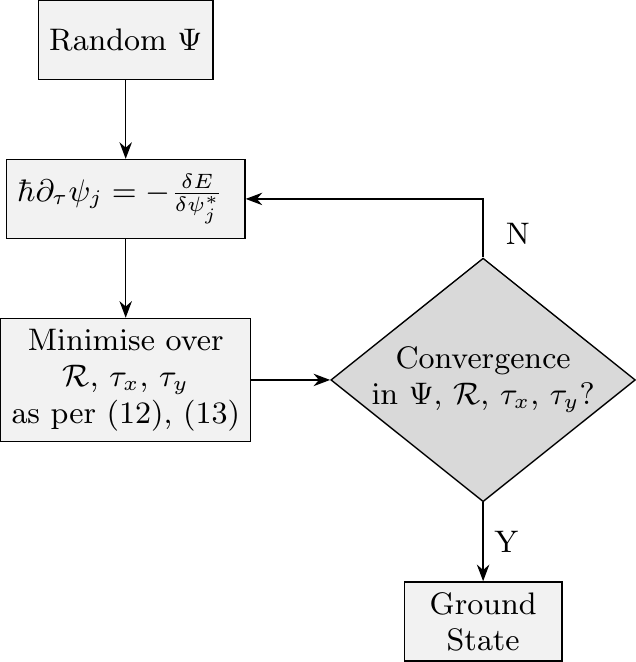}
	\caption{Schematic description of the algorithmic procedure. The equation of motion in the top left of the figure is obtained by Wick-rotating the Gross-Pitaevskii equation to imaginary time $\tau=it$.}
	\label{fig:flowchart}
\end{figure}

The minimisation of \eqref{Ed} can then be performed numerically by repeatedly alternating the minimisation with respect to $\Psi$, $\mathcal{R}$ and $\tau_x$, $\tau_y$. As in \cite{Noi}, the minimisation over $\Psi$ is performed by solving the imaginary-time Gross-Pitaevskii equation using a split-step method. In practice, we find that it is most efficient to perform more steps to evolve $\Psi$ and less for the remaining parameters. A schematic description of this algorithmic procedure is given in the following Fig.~\ref{fig:flowchart}. Furthermore, for the highly symmetric vortex lattices we find, the latter parameters converge to simple values (e.g. $\mathcal{R}=\sqrt{3}$). Finally, one must check for convergence in the time step and the discretisation lattice constants.

\section{Results}
\label{Sec:4}
In the following, we will be concerned with the case of species of equal masses $m_1=m_2$ and equal particle density in the repulsive interaction regime $g_{12}\ge 0$. Although our method can treat the attractive regime as well, its solutions for the equal masses are simple: the ground state solution will consist of two perfectly overlapping triangular lattices.
The scenario involving different mass ratios (though in a harmonic trap) has been considered in \cite{Kuopanportti12}. 
An early  important result for equal masses in the repulsive regime was obtained semi-analytically in the LLL \cite{Mueller02}, assuming equal scattering lengths for the two coupled systems $a_{11}=a_{22}$ (and consequently equal intra-species interactions for this equal masses case). This assumption in particular allows for the achievement of an SU(2) symmetric system. One consequence for such a system is, for example, that the system becomes invariant under the exchange of the two superfluids.

In \cite{Mueller02}, the two superfluids were found to transition, at the variation of the parameter $\alpha=g_{12}/\sqrt{g_1g_2}$, between four different states. 
At low interacting strengths ($0<\alpha<T_1=0.172$) the ground state consists of two interlaced triangular lattices with a vortex of the first species centered between three vortices of the second species. At $T_1$ the first transition occurs: for $T_1<\alpha<T_2=0.373$ the system is found to be made of two interlaced oblique lattices with varying angle $\phi$; the vortices of the first species are now sitting centered between four vortices of the second species. The second transitions occurs when $\phi=90^{\circ}$ giving place to two interlaced square lattices. The system remains stable in this state for $T_2<\alpha<T_3=0.926$ until the third and last transition takes place. For $\alpha>T_3$ the square lattices, following a spontaneous breaking of symmetry, continuously stretches into interlaced rectangular lattices of aspect ratio ${R}$. We recall that above the boundary $\alpha=1$ the two superfluids become immiscible and      the so-called stripe phase is obtained \cite{Cooper08}. In this region the density of each superfluid concentrates in the central area perpendicular to the long side of the rectangle.  
\begin{figure}[h!]
	\centering
	\newlength\figureheight 
	\newlength\figurewidth 
\includegraphics{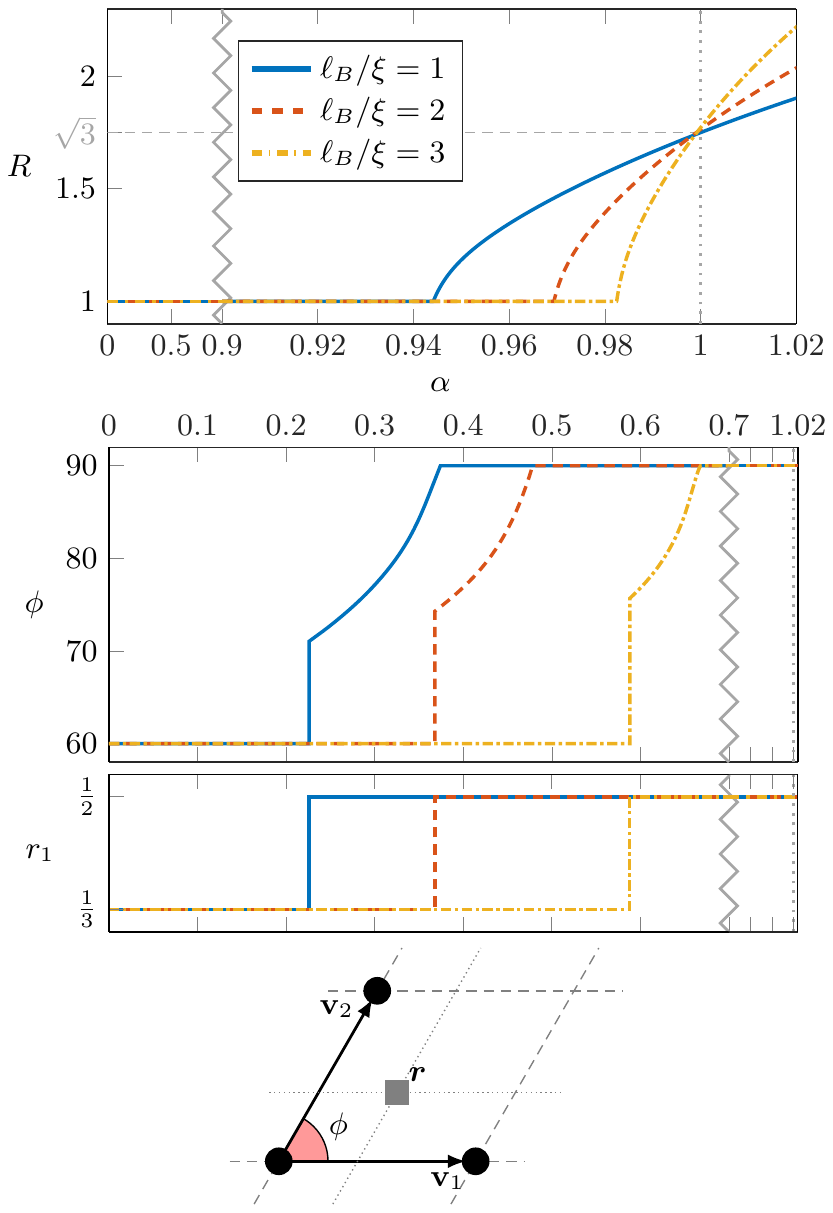}   
	\caption{Extension of the results from \cite{Mueller02}. When varying $\alpha$, the parameter ${R}$ describes the second order transition  transforming a square lattice into a rectangular lattice; the parameter $\phi$ instead, experiences at first a jump, signalling a first order transition responsible for the transformation of the triangular lattice into the oblique lattice. Further observing the behaviour of $\phi$, it is possible to spot where another second order transition occurs, continuously transforming the oblique lattice into the square lattice. The last diagram defines the parameters $R=\frac{|\mathbf{v}_2|}{|\mathbf{v}_1|}$ and $\phi=\arccos{(\mathbf{\hat{v}}_1\cdot \mathbf{\hat{v}}_2)}$. The vector ${\boldsymbol{r}=r_1{\mathbf{v}}_1+r_2{\mathbf{v}}_2}$ defining the relative translation between the two species can be expressed in terms of the parameters $\tau_x$ and $\tau_y$, as explained in Appendix, by appropriate coordinate transformations. 
	For components of equal masses one always obtains minimisers satisfying ${r_1=r_2}$. At the occurrence of the first order transition $r_1$ experiences a discontinuity as well: this permits the transition from the triangular to the square configurations.}
	\label{fig:afew}
\end{figure}
Because these results were obtained in the LLL, the connection to experiments is not immediate as most experiments on vortex lattices are away from this regime. The method outlined in the previous sections allows for the extension of these results to regimes of larger intraspecies-interaction or slower rotation rates.

\begin{figure}[h!]
	\centering
	\includegraphics[scale=1]{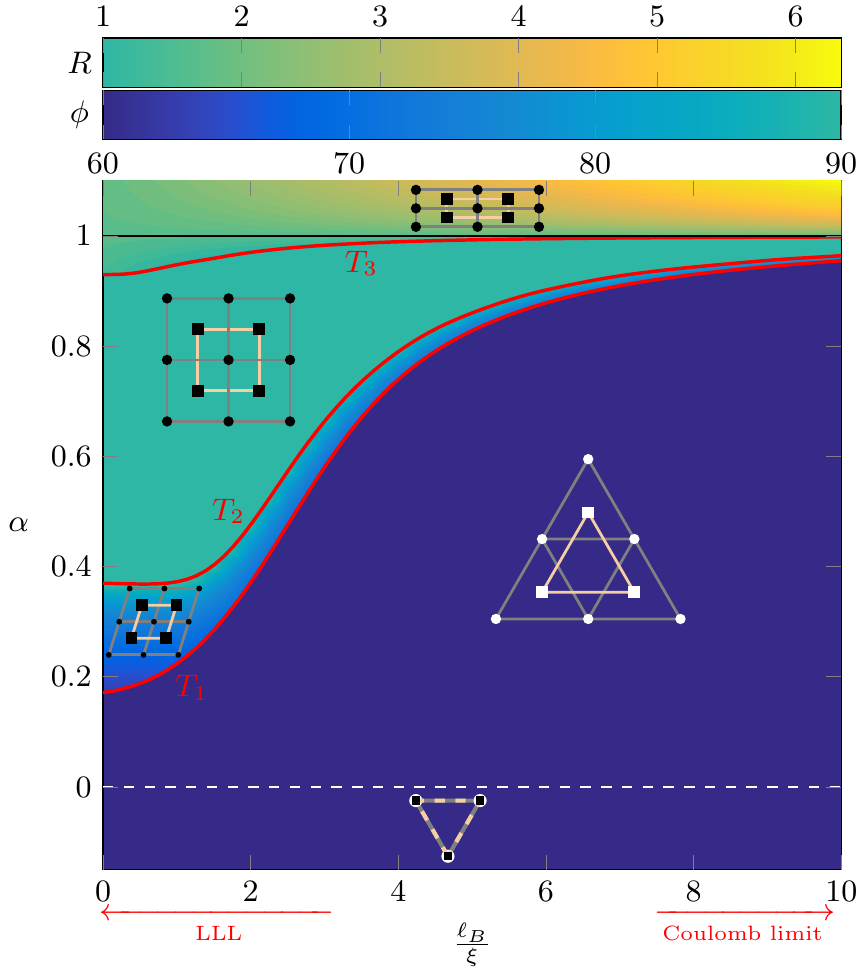}	
	\caption{Phase space describing the ground states of two interacting superfluids of equal masses and same particle number per unit cell ${\mathcal{N}_1=\mathcal{N}_2}$. The abscissa represents the intraspecies interaction strength (which is assumed to be the same for both species) while the  ordinate the interaction strength amongst the two different species. The area of the phase space below $T_2$ is  characterised by the parameter $\phi$, while that above is characterised by ${R}$.  Transition $T_1$ is of first order, while $T_2$ and $T_3$ are second order transitions. For completeness the trivial attractive regime ($\alpha<0$) is included as well, showing a ground state consisting of two overlapping triangular lattices.}
   \label{fig:phasespace}
\end{figure}
In Fig.~\ref{fig:afew} we present a detailed characterisation of the transitions undergone by
the system as reflected by the behaviour of the  the two parameters $\phi$ 
and ${R}$ (see also  Fig.\  \ref{fig:phasespace}).
In particular, $\phi$ experiences a discontinuous jump at $T_1$ and has a discontinuous derivative at $T_2$.
On the other hand, ${R}$ has a discontinuous derivative at $T_3$.
This result can be directly compared with that of \cite{Mueller02}. It is also possible to notice that, at the $SU(2)$ symmetric point, the lattice configuration is independent of the strength of the interactions. Here we find a lattice configuration consisting of two interlaced rectangular lattices of aspect ratio ${R}=\sqrt{3}$, such that the combination of the two lattices gives rise to a triangular lattice.

It is now possible to go even further and explore the phase diagram going towards the Coulomb limit:  Fig.~\ref{fig:phasespace} shows the complete phase diagram for the ground states of two interacting superfluids. As can be intuitively expected, for $\alpha<0$ a configuration consisting of two non interacting triangular lattices is found: the ground state is degenerate with respect to translations of the two lattices. 
In the particular case of $\alpha=0$, a configuration consisting of two non-interacting triangular lattices is found.   The ground state is degenerate with respect to translations of the two lattices. The red lines in  Fig.~\ref{fig:phasespace} mark the three phase boundaries $T_1$, $T_2$ and $T_3$ corresponding to each phase transition; the colours encode the value of either $\phi$ or ${R}$. For states below $T_2$ the only varying parameter is $\phi$. The colour coding the highest value of $\phi$ is the same as the colour coding the lowest value of ${R}$: this appears in the region between $T_3$ and $T_2$, where neither of these two parameters varies. Above $T_3$ the varying parameter is ${R}$ and the colour code changes accordingly. 
Although in the LLL the square configuration is predominant, our results demonstrate that in the Coulomb limit the triangular lattice configuration takes over while the other configurations are suppressed.

\begin{figure}[h!]
	\centering
	\setlength\figureheight{\linewidth} 
	\setlength\figurewidth{\linewidth}
\includegraphics{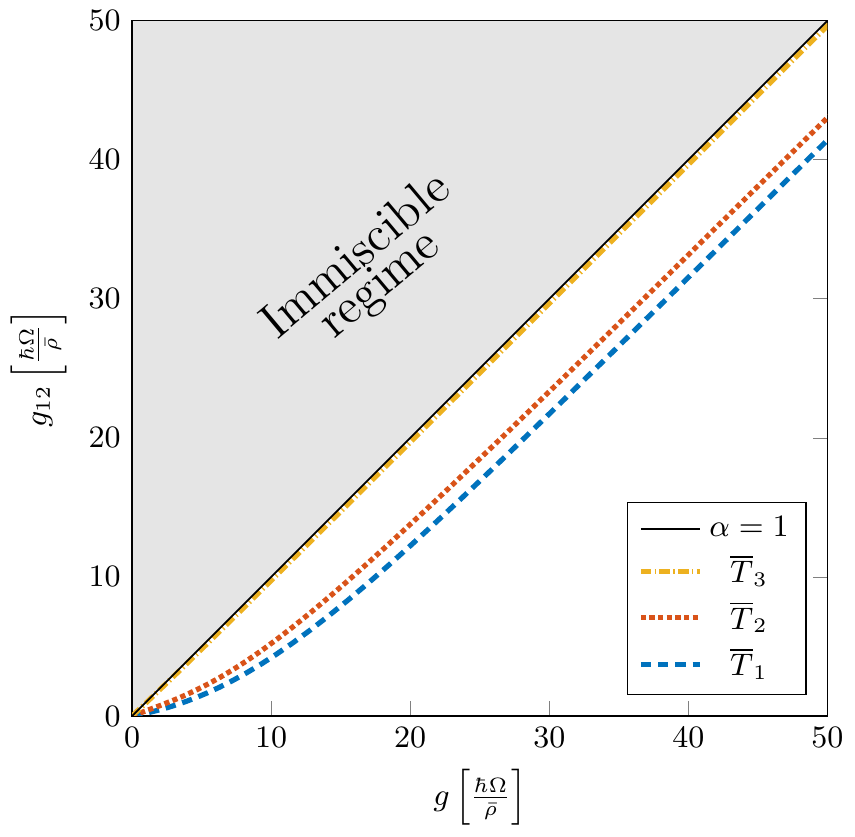}	
	\caption{Linear phase boundaries in the miscible regime. $\overline{T}_1$ marks the boundary between the triangular phase and the oblique phase, $\overline{T}_2$ divides the oblique and the square phases and $\overline{T}_3$ is the last phase boundary leading to the rectangular phase.}
	\label{fig:lineartrans}
\end{figure}

While it is convenient to study the phase space in Fig.~\ref{fig:phasespace} as a function of the parameters $\alpha$ and $\ell_B/\xi$, this approach conceals some very simple properties of the phase boundaries $T_1$, $T_2$ and $T_3$. In Fig.~\ref{fig:lineartrans} the phase
diagram is plotted in terms of the alternative parameters
$g$ and $g_{12}$. One sees that the phase boundaries asymptotically become linear in the Coulomb regime. 
An argument explaining this behaviour goes as follows.
Deep in the Coulomb regime, the energy of the system is dominated by terms representing interactions.  
In this limit, one can write the energy density as 
$
{\mathcal{E}(g,g_{12})\sim \frac{g}{2}(\rho_1^2+\rho_2^2)+g_{12}\rho_1\rho_2}=\frac{1}{2} g \rho^2 + (g_{12}-g) \rho_1 \rho_2
$
where $\rho=\rho_1+\rho_2$ is the total density.    
Since a phase boundary $\overline{T}(g)$ between a phase configuration $A$ and a configuration $B$ can be defined as the value of the interspecies strength such that ${E^{A}(g,g_{12}=\overline{T})=E^{B}(g,g_{12}=\overline{T})}$, it is possible to write an expression for $\overline{T}(g)$. In particular, one finds that
\begin{align}
\frac{\overline{T}(g)}{g}=\frac{1}{2} \frac{\langle \rho_{A,1}^2\rangle+\langle \rho_{A,2}^2 \rangle-
\langle \rho_{B,1}^2\rangle-\langle \rho_{B,2}^2\rangle}{\langle \rho_{B,1}\rho_{B,2} \rangle - \langle \rho_{A,1}\rho_{A,2}\rangle}
\end{align}
where brackets denote spatial average.
Next, we note that deep in the Coulomb regime, variations in the \emph{total}  density are energetically prohibitive 
and so the total density, at this level of approximation, is constant. 
For instance, while $\rho_1$ will approach zero near a vortex in $\psi_1$, $\rho_2$ will have a local maximum there, making 
the total density nearly constant.   Writing   density with respect to its average as $\delta \rho_{A,1} = \rho_{A,1} - \langle \rho_{A,1} \rangle$ (with similar notation for
the other components) we then have

\bes{
\frac{\overline{T}(g)}{g} &=\frac{\langle \delta \rho_{A}^2\rangle
-2 \langle \delta \rho_{A,1}\delta \rho_{A,2} \rangle
-\langle \delta \rho_{B}^2 \rangle
+2 \langle \delta \rho_{B,1}\delta \rho_{B,2} \rangle}
{2\left(\langle {\delta}\rho_{B,1}{\delta}\rho_{B,2} \rangle - \langle \delta \rho_{A,1} \delta \rho_{A,2}\rangle\right)}.
}{}
In the Coulomb limit, the variances in the total densities become negligible and the leading order behaviour of the phase boundaries can be found to be
\bes{
\lim_{g\rightarrow\infty}\frac{\overline{T}(g)}{g}=1.
}{scalingCoul}
Therefore, the phase boundaries have the form ${\overline{T}=g+a_j}$ 
where the intercepts $a_j$, are determined by the kinetic energy difference between the two configurations and likely
cannot be determined from such simple arguments.  
Operating the appropriate transformations to the phase boundaries in Fig.~\ref{fig:phasespace}, we obtain the linear phase boundaries $\overline{T}_i$ presented in Fig.~\ref{fig:lineartrans}.  The numerical solution for the phase diagram indeed verifies these
simple arguments.

For computational convenience, the phase space in Fig.~\ref{fig:phasespace} was calculated with two vortices per species per unit cell. Considering a unit cell containing only one vortex per species, 
as done for instance in the early work by Abrikosov \cite{Abrikosov57}, does not allow for configurations other than the square and rectangular lattices. The smallest unit cell needed to obtain the correct ground states contains a minimum of two vortices (per species).
The results obtained in this setting can be found to be consistent with those obtained in larger unit cells, as long as the  size of the cell is appropriate (namely if the unit cell contains an even number of vortices). For other unit cells (e.g.~a unit cell containing an odd number of vortices per species) one will in general observe frustrated lattices. However, such configurations of the system have higher energy densities and are therefore disregarded.

\section{Conclusions}
In conclusion, we have presented an extension of the method outlined in \cite{Noi} to treat multicomponent systems. The addition of each new component to the system must be complemented with the introduction of two new phase factors accounting for relative translations. The energy functional must then be minimised over these parameters as well as over the aspect ratio and the wavefunction. In particular, we have shown it is possible to find an exact expression for the minimisers of the energy functional when $\Psi$ is held fixed. Under this new framework, it has been possible to obtain an extension of the results for scalar multicomponent superfluids, until now limited to the LLL, to strongly interacting systems (Coulomb limit). In particular, we have shown the results obtained in the LLL do not extend to the Coulomb limit where a triangular lattice configuration is found to dominate the phase diagram. Nonetheless, the lattice configuration at the $SU(2)$ symmetric point remains invariant with respect to changes in the interaction strengths. Finally, from simple general considerations on the energetics of the coupled system, we have shown that, in the Coulomb limit, the phase boundaries can be described by a linear relation. This in turn, also provides an explanation as to why in the Coulomb limit the triangular phase is dominant. It can be in fact deduced from Fig.~\ref{fig:phasespace}, that the transitions $T_j$ each go to one deep in the Coulomb limit, as can in turn be inferred from the scaling argument presented in \eqref{scalingCoul}.

The extended framework employed to obtain such results can be directly applied to study systems where the mass ratio differs from unity where richer lattice configurations are expected. Another intriguing application is that of a system under more general synthetic gauge fields. 
\begin{acknowledgements}
This work was supported in part by the European Union's Seventh Framework Programme for
research, technological development, and demonstration
under Grant No.\ PCIG-GA-2013-631002.
\end{acknowledgements}
\appendix
\section{Computational Framework Details}
\label{apptrans}

In this Appendix we will describe the details of the generalisation of the computational framework
presented in \cite{Noi} to multi-component systems.  The subtle aspect of such a generalisation
corresponds to the phase factors
$e^{-i \tau_ja_j}$ entering \eqref{Wphimatrix}.  In \cite{Noi} these phase factors were set to unity.

To begin, for simplicity, we will focus on a single-component continuum BEC under uniform rotation.
The physical quantities describing such a system are the gauge invariant velocity
$\boldsymbol{v}= \frac{\hbar}{m} \nabla \theta - \frac{1}{m} {\bf A}$ and the superfluid density $\rho$.  Both of these quantities
follow from the condensate order parameter $\psi = \sqrt{\rho} e^{i\theta}$ and the vector potential 
corresponding to uniform rotation for which we choose the Landau gauge: ${\bf A} = 2m \Omega (0,x)$.

Now let us consider an infinite periodic vortex lattice.  Without loss of generality, we may choose an
$L_x \times L_y$ rectangular unit cell that tiles the system.  The superfluid velocity and density must have
the periodicity of this unit cell.  In particular, by integrating the equations ${\boldsymbol{v} (x+L_x,y)=\boldsymbol{v}(x,y+L_y)= \boldsymbol{v}(x,y)}$
one finds that the phase must satisfy
\begin{align}
\theta(x+L_x,y) &= \theta(x,y) + \frac{2\Omega m}{\hbar}  L_x y + \kappa_x    \\
\theta(x,y+L_y) &= \theta(x,y) +\kappa_y 
\label{periodicity}
\end{align}
where $\kappa_x$ and $\kappa_y$ are constants of integration. For the sake of convenience and clarity, let us introduce a rescaled version of these constants, namely the phases $\tau_j=\kappa_j/L_j$ appearing above in \eqref{phgt} and \eqref{phgt2}.
Next we introduce the magnetic translation operator ${{\cal T}(\boldsymbol{r}) = e^{\frac{i}{\hbar}{\mathbf{\Pi} \cdot \boldsymbol{r}}}}$, 
where ${\Pi_x = p_x -2m\Omega y}$ and ${\Pi_y = p_y}$ are the generators of magnetic translation in the Landau gauge \cite{Noi}.  Then one can verify
that the periodicity condition for the superfluid density, $\rho(x,y) = \rho(x+L_x,y)=\rho(x,y+L_y)$ and velocity,  Eq.\ (\ref{periodicity}),
can be written succinctly as
\bes{
e^{\frac{i}{\hbar}\Pi_x L_x} \psi(x,y) &= e^{i \tau_xL_x} \psi(x,y), \\
e^{\frac{i}{\hbar} \Pi_y L_y} \psi(x,y) &= e^{i \tau_yL_y} \psi(x,y).
}{mtgbc}

Now let us consider magnetically translating this wave function by $-\boldsymbol{r}$ where 
${\boldsymbol{r} = (r_x,r_y)}$:  ${\tilde{\psi}(x,y) \equiv {\cal T}(\boldsymbol{r}) \psi(x,y)}$.  
Due to the symmetries of the problem (namely that the generators of magnetic translation
commute with the kinetic momenta), the 
energy per unit area corresponding to $\psi(x,y)$ is the same as that of  $\tilde{\psi}(x,y)$.  
Moreover the densities of these two wave functions are identical apart from translation:
$\tilde{\rho}(x,y) \equiv |\tilde{\psi}(x,y)|^2 = \rho(x+r_x,y+r_y)$.
Therefore the vortex lattice given by $\psi$ is related to that given by $\tilde{\psi}$ by a simple translation.
By choosing $\boldsymbol{r}$ to satisfy 
$2m\Omega r_y= -\hbar\tau_x$
and
$2m\Omega r_x= \hbar\tau_y$
we have the simplified boundary condition
\begin{align}
e^{\frac{i}{\hbar}\Pi_x L_x} \tilde{\psi}(x,y) &= \tilde{\psi}(x,y) \\
e^{\frac{i}{\hbar} \Pi_y L_y} \tilde{\psi}(x,y) &=  \tilde{\psi}(x,y)
\end{align}
which was the condition taken by us previously in \cite{Noi}.
A closer look 
 at \eqref{mtgbc}  reveals that one can alternatively consider the following transformation of the operators of the magnetic translation group
\bes{
\Pi_j\rightarrow\Pi_j-\hbar\tau_j.
}{}
This corresponds to a gauge transformation $\psi \rightarrow e^{i\lambda} \psi$ with  $\lambda=\tau_xx+\tau_yy$.  
As can be readily verified from \eqref{mtgbc}, the transformed wave function is invariant under magnetic translation across a unit cell. 
Finally, the terms $a_j$ entering the phase factors $e^{-i\tau_ja_j}$ in \eqref{Wphimatrix} arise from the Peierls integrals calculated over the Hofstadter computational lattice vectors.
 
Through the above considerations, one sees that by specifying $\tau_x$ and $\tau_y$, a particular unit cell of the vortex lattice is
specified.  Changing $\tau_x$ and $\tau_y$ will translate this unit cell, but will not affect the energy per unit cell or the vortex geometry
of the periodic system.  Thus, without loss of generality, we can set $\tau_x=\tau_y=0$ for the single component system.  However,
for the two-component system, such a freedom does not exist.  In the method described Sec.\ \ref{Sec:3}, we have set the 
$\tau$-parameters for the first component to zero, while keeping those of the second component as degrees of freedom to be minimised over.

\bibliographystyle{apsrev4-1}

%

\end{document}